%% file: nlurkin_moriond.tex
\documentclass{moriond}

\pdfoutput=1
\usepackage[alsoload=hep,binary-units=true]{siunitx}
\usepackage{subfig}
\usepackage{amsmath}
\usepackage{xspace}

\DeclareSIUnit\eVperc{\eV\per\clight}
\DeclareSIUnit\gev{\giga\eV}
\DeclareSIUnit\mev{\mega\eV}
\DeclareSIUnit\gevperc{\giga\eV\per\clight}
\DeclareSIUnit\mevperc{\mega\eV\per\clight}
\DeclareSIUnit\gevpercsq{\giga\eV\per\clight\squared}
\DeclareSIUnit\mevpercsq{\mega\eV\per\clight\squared}
\DeclareSIUnit{\nothing}{\relax}

\def\Journal#1#2#3#4{{#1} {\bf #2}, #3 (#4)}

\def\NIMA{{\em Nucl. Instrum. Methods} A}

\def\PLB{{\em Phys. Lett.}  B}
\def\PRL{\em Phys. Rev. Lett.}
\def\PRD{{\em Phys. Rev.} D}

\def\EPJC{{\em Eur. Phys. J.} C}
\def\JHEP{{\em JHEP}}

\def\be{\begin{equation}}
\def\ee{\end{equation}}
\def\bea{\begin{eqnarray}}
\def\eea{\end{eqnarray}}

\newcommand{\kpmunu}{K^+\to\mu^+\nu}
\newcommand{\kpmuN}{K^+\to\mu^+N}

\newcommand{\Photo}{}

\begin{document}
\input{cfg_commands} 

\vspace*{4cm}
\title{Heavy neutrino searches and NA62 status}

\author{ N. Lurkin \footnote{for the NA62 collaboration: 
G.~Aglieri Rinella, R.~Aliberti, F.~Ambrosino, R.~Ammendola, B.~Angelucci, 
A.~Antonelli, G.~Anzivino, R.~Arcidiacono, I.~Azhinenko
% ...
}}

\address{School of Physics and Astronomy, University of Birmingham, \\
B15 2TT, Birmingham, United Kingdom}

\maketitle

\abstracts{
The NA62 experiment at CERN SPS recorded in 2007 a large sample of \(\kpmunu_\mu\) decays.
A peak search in the missing mass spectrum of this decay is performed. 
In the absence of observed signal, the limits obtained on \(\br{K^+\to\mu^+\nu_h}\) and on the mixing matrix element \(\abs{U_{\mu 4}}^2\) are reported. 
The upgraded NA62 experiment started data taking in 2015, with the aim of measuring the branching fraction of the \(K^+\to\pi^+\nu\bar{\nu}\) decay.
An update on the status of the experiment
is presented.
}

\section{Heavy neutrino searches in \texorpdfstring{\(K^+\to\mu^+\nu_\mu\)}{K+mu+numu} decays}
\labelsec{hnu-searches}
With the increasing evidences that neutrinos have non-zero masses, it is necessary to extend the Standard Model (SM) to accommodate them.
The Neutrino Minimal Standard Model (\(\nu\)MSM)~\cite{asaka_2005} solves this problem by adding 3 massive right-handed neutrinos to the SM.
Effective vertices with the \(W^\pm, Z\) and the SM leptons can be built, with a mixing matrix \(U\) describing the mixing between the heavy neutrinos and the SM neutrinos.
The SM neutrinos then acquire masses through the see-saw mechanism.
For heavy neutrinos with masses below the kaon mass, limits on their mixing matrix elements can be placed by searching for peaks in the missing mass spectrum of \(K^\pm\) decays \cite{shrock_1980}.

This analysis focuses on the \(\kpmunu_\mu\) decay to search for heavy neutrino in the mass range \SIrange{300}{375}{\mevpercsq}.
Strong limits of the order of \num{e-8} (up to \SI{300}{\mevpercsq}) and \num{e-6} (up to \SI{330}{\mevpercsq}) on the element \(\abs{U_{\mu 4}}^2\) are already set by stopped kaon experiments \cite{artamonov_2015,hayano_1982}.
The heavy neutrino acceptance drops quickly above \SI{375}{\mevpercsq}.
The following assumptions are made: the heavy neutrinos decay only to SM particles, and \(\abs{U_{\mu4}}^2 < \num{e-4}\), such that the mean free path of heavy neutrinos in the mass range considered is longer than \SI{10}{\km}.
Their decay can then be neglected as the probability of decaying in the detector of decay volume is below \perc{1}.
Because only the production process is looked at, the limits scale linearly with the kaon flux.

\subsection{The NA62 setup in 2007}
During the data-taking campaign of 2007, the NA62 experiment collected a large sample of kaon decay in-flight.
The high-efficiency single track trigger was designed to collect \(\kenu_e\) and \(\kmunu_\mu\) decays, aiming at a test of the lepton universality in the kaon decay \cite{lazzeroni_2013}.
Part of the \(\kmunu_\mu\) sample recorded is used to search for the production of heavy neutrino.

The beam line, described in detail in Batley {\it et al.}~\cite{batley_2007}, was designed to provide simultaneous \(K^+\) and \(K^-\) beams. 
They were extracted from the \SI{400}{\giga\eVperc} SPS proton beam impinging on a \SI{40}{\cm} long beryllium target. 
The final beam momentum of \SI{74.0(14)}{\giga\eVperc} was selected using a system of dipole magnets and a momentum-defining slit incorporated into a beam dump. 
This \SI{3.2}{\m} thick copper/iron block provided the possibility to block either of the \(K^+\) or \(K^-\) beams. 
The beams were focused and collimated before entering the \SI{114}{\m} long cylindrical vacuum tank containing the fiducial decay volume. 
The beam contained mainly pions but included approximately \SI{6}{\percent} of kaons.
For about \num{1.8e12} primary protons incident on the target per SPS pulse of \SI{4.8}{\s} duration, the secondary beam flux at the entrance to the decay volume was \num{2.5e7} particles per pulse.

The momenta of the charged particles were measured by a magnetic spectrometer housed in a tank filled with helium at approximately atmospheric pressure.
It was composed of four drift chambers (DCHs) and a dipole magnet located between the second and third chambers. 
The magnet provided a horizontal transverse momentum kick of \SI{265}{\mega\eVperc}, and had a resolution of \(\sigma_p/p = \perc{0.48}\oplus\perc{0.009}\cdot p\) where \(p\) is the particle momenta in \SI{}{\gevperc}.
A hodoscope (HOD) composed of two planes of plastic scintillator was placed after the spectrometer to provide precise timing of the charged particles and generate fast trigger signals for the low-level trigger. 
A \SI{127}{\cm} thick quasi-homogeneous electromagnetic calorimeter filled with liquid krypton (LKr) was located downstream. 
The volume is divided into \num{13248} cells of \(\sim 2\times 2\SI{}{\;\cm\squared}\) cross section without longitudinal segmentation. 
The energy resolution is \(\sigma_E/E = \perc{3.2}/\sqrt{E} \oplus \perc{9}/E \oplus \perc{0.42}\) and the position resolution is \(\sigma_x = \sigma_y = \SI{4.2}{\mm}/\sqrt{E} \oplus \SI{0.6}{\mm}\) where the particle energy \(E\) is given in \si{\gev}.
A muon veto system (MUV) was installed behind the LKr and consisted of three planes of scintillator orthogonal to the beam axis, each one preceded by a \SI{80}{\cm} thick iron wall. 
They were made of \SI{2.7}{\m} long and \SI{2}{\cm} thick strips alternatively arranged horizontally and vertically. 
The width of the strips was \SI{25}{\cm} in the first two planes and \SI{45}{\cm} in the last one. 
More details on the detector can be found in Fanti {\em et al.}~\cite{fanti_2007}.

The main trigger condition for selecting the \(K^+\to\mu^+ N\) decays (\(N=\nu_\mu,\nu_h\)) required at least one coincidence of hits in the two HOD planes, and bounds on the hits multiplicity in the DCH.
This trigger line was downscaled by a factor 150. 
To obtain the purest sample, only the data taking periods with single \(K^+\) beam (\perc{43} of the integrated kaon flux, as in Lazzeroni {\it et al.}~\cite{lazzeroni_2011}) are used to set the limits. 
Periods with single \(K^-\) beam, where the beam halo background is higher, are used to study the background from beam halo muons.
A horizontal lead bar was installed between the two HOD planes, for muon identification studies, during periods with simultaneous beam.
Because this lowered the vetoing power of the LKr calorimeter for photons, these periods are not used at all for this analysis. 

\subsection{Analysis strategy}
\labelsec{nuh_strategy}
In the decay \(\kpmuN\), the neutrino missing mass can be reconstructed as \(m_h = \sqrt{m^2_\text{miss}} = \sqrt{(p_K-p_\mu)^2}\).
The kaon four-momentum \(p_K\) is the nominal beam kaon one, measured from \(\kpipipi\) samples at regular interval (\(\sim 500\) bursts).
The muon four-momentum \(p_\mu\) is that of a charged track reconstructed in the spectrometer, under the assumption that it is a muon.
A detailed GEANT3 simulation of the detector is used to perform Monte Carlo (MC) simulations of the signal and background channels. 

The acceptance and resolution of the \(\kpmunu_h\) channel are studied from a set of MC simulations.
The heavy neutrino mass was varied between \SI{240} and \SI{380}{\mevpercsq} at \SI{1}{\mevpercsq} intervals.
The \(\kpmunu_\mu\) and other kaon decay background channels are also simulated to determine the expected spectrum of the reconstructed \(m_\text{miss}\) spectrum.
The contribution from the beam halo is evaluated using a control data sample, defined as the sample recorded with the \(K^-\) beam only.

The observed and expected \(m_\text{miss}\) spectra are compared to set limits on the number of observed \(\kpmunu_h\) decays for each tested \(m_h\).
These limits are translated into limits on the heavy neutrino production branching ratio \(\br{\kpmunu_h}\), and limits on the mixing matrix element \(\abs{U_{\mu 4}^2}\). 

\subsection{Event selection}
The event selection requires a single positively charged track, which falls in the geometrical acceptance of the DCH, LKr and MUV detectors.
The momentum must be between \SI{10}{\gevperc} and \SI{65}{\gevperc} and in-time with the HOD trigger time.
The closest distance of approach (CDA) between the track and the nominal beam direction must be smaller than \SI{3}{\cm}.
In order to reject non-muon tracks, the track extrapolated position on the MUV detector must match MUV signals from the first two planes, in time and space.

Because the signal events does not feature any electromagnetic energy, it is required that no cluster of energy deposition above \SI{2}{\gev}, and not associated with the track, is present in the LKr.
Clusters are considered to be associated with a track if they are consistent with bremsstrahlung emission from the track upstream: within \SI{6}{\cm} from the straight-line extrapolated upstream track; 
or bremstrahlung emission from the track downstream of the spectrometer: within \SI{40}{\cm} from the extrapolated track impact point.

To suppress the beam halo muons component, five-dimensional cuts in the \((z_\text{vtx},\theta,p,\text{CDA},\phi)\) space are applied, where \(z_\text{vtx}\) is the longitudinal position of the reconstructed vertex, \(\theta\) is the angle between the \(K^\pm\) and \(\mu^\pm\) direction, and \(\phi\) is the azimuthal angle of the muon in the transverse plane.
As mentioned earlier, the signal region is restricted to \(\SI{300}{\mevpercsq}<m_\text{miss}<\SI{375}{\mevpercsq}\).

\subsection{Background estimation}
The total number of kaon decays in the fiducial volume is \(N_K = \num{5.977e7}\), taking into account the trigger downscaling of 150.
The distribution of background from kaon decays, shown in \reffig{stack_m}, is extracted from the simulation and scaled using \(N_K\).
The \(\kpmunu_\mu\) channel forms a peak at zero \(m^2_\text{miss}\), whose width is determined by the kaon momentum spectrum, the beam divergence and the spectrometer resolution.
This component is well outside the signal region and only the radiative tail from \(\kpmunu_\mu\gamma\) must be taken into account.
In the signal region, the dominant background comes from the \(K^+\to\pi^0\mu^+\nu_\mu\) decay with undetected \(\pi^0\) due to the non-hermetic geometrical acceptance.
The \(K^+\to\pi^+\pi^+\pi^-\) and \(K^+\to\pi^+\pi^0\pi^0\) decays must be taken into account for the same reason, but are highly suppressed due to the presence of either three tracks, or multiple photons.

\begin{figure}[htb]
	\centering
	\subfloat[\labelfig{stack_m}]{\includegraphics[width=.5\textwidth]{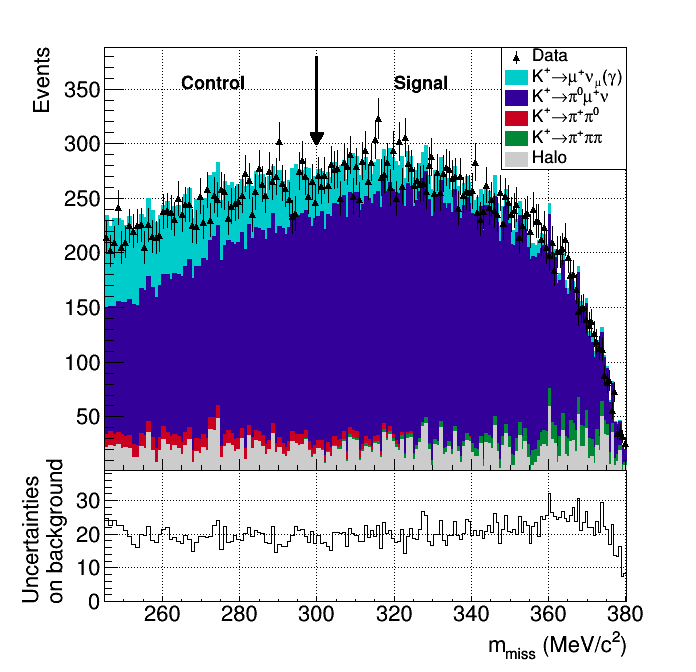}}
	\subfloat[\labelfig{sqerr}]{\includegraphics[width=.47\textwidth]{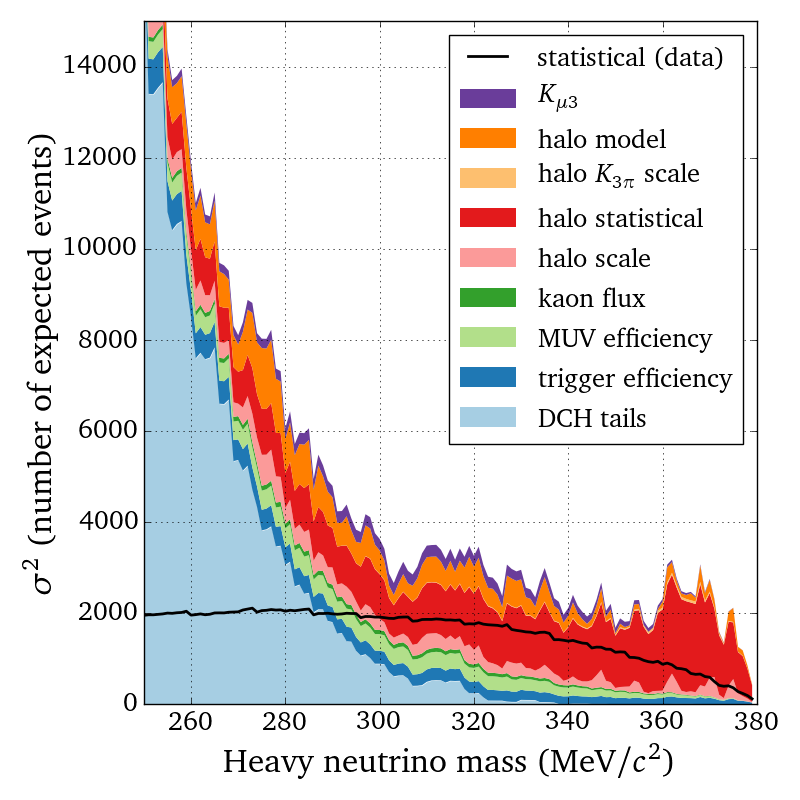}}
	\caption{
		(a) Missing mass distribution of data and background estimate in the signal and control regions.
		Error bars are data statistical errors.
		The lower plot shows the total uncertainty on the background estimate.
		(b) Squared uncertainties on the number of expected background events at each heavy neutrino mass.
		The coloured bands are the different contributions to the systematic uncertainties.
		The black line is the statistical contribution.
	}
\end{figure}

The muon halo background is extensively studied with the control sample recorded during the \(K^-\) beam period, resulting in the five-dimensional cut mentioned in \refsec{nuh_strategy}.
The residual halo background is modelled and normalized using the control sample.
A contamination from \(K^-\to\pi^-\pi^-\pi^+\) (\(\kppp\)) decays, where the pion is mis-identified as a muon, must be subtracted from the control sample.
The normalization is computed using the halo dominated region obtained by requiring \(3 < \text{CDA} < \SI{8}{\cm}\).
Although the contribution itself, as seen in \reffig{stack_m} is relatively small, it dominates the total uncertainty on the background estimation.
Multiple systematic uncertainties related to the halo background are taken into account.
They are associated to the limited size of the control sample, the subtraction of the \(\kppp\) component, the normalization from the control sample to the final sample, and the assumption that the halo model in the control sample accurately reproduces that of the \(K^+\) data.
The uncertainty on the normalization is assessed by normalizing the halo contribution using an alternative halo-dominated region at negative \(m^2_\text{miss}\).
The halo model contribution is evaluated by comparing the final kinematic distributions of halo events obtained from the control sample, and from a sample recorded during a period where both beams were blocked.

The muon identification and the detector resolution are also studied on data with a \(\kpmunu_\mu\) sample and a \(K^+\to\pi^+\pi^0 (\pi^0\to\gamma\gamma\)) sample reconstructed using LKr only, respectively.
Their contribution to the uncertainty is small in the signal region.

The trigger efficiency depends only on the number of tracks, and being the same for the signal and \(\kpmunu_\mu\) decays, it cancels out to a very good approximation. 
The trigger efficiency for the background is different due to possible additional electromagnetic activity in the detectors.
It is evaluated in the region \(245 < m_\text{miss} < \SI{298}{\mevpercsq}\) since strong limits on the heavy neutrino production already exist in this range.
Because the background trigger efficiency does not depend on the missing mass, and the background composition is similar in \(245 < m_\text{miss} < \SI{375}{\mevpercsq}\), it is taken as \(\group{89.8\pm0.5_\text{stat}\pm0.4_\text{syst}}\%\) in the whole range.  

All the contributions to the systematic uncertainties can be seen in \reffig{sqerr}.
The total uncertainties on the expected background (systematic and statistical) are displayed in the lower plot of \reffig{stack_m}.

\subsection{Limits on heavy neutrino production}
A peak search of the \(m_\text{miss}\) distribution in steps of \SI{1}{\mevpercsq} is performed.
For each heavy neutrino mass \(m_h\) considered, a window of size \(\pm\sigma_h = \SI{120}{\mevpercsq}-0.03\cdot m_h\) is used, corresponding to the resolution on the heavy neutrino mass.
The upper limits at \perc{90} confidence level (CL) on the number of reconstructed \(\kpmunu_h\) events \(n_{UL}\) are computed by applying the Rolke-Lopez method~\cite{rolke_2001} for the case of a Poisson process in presence of gaussian background.
The input necessary to the computation are the number of observed data events, the number of expected background events, and their uncertainties.

The local significance never exceed \(3\sigma\) and therefore, no signal is observed.
The obtained limits are shown in \reffig{mu}.
These upper limits are translated into upper limits on the heavy neutrino branching ratio, and on the mixing matrix element \(\abs{U_{\mu 4}}^2\) using the relations
\begin{alignat*}{2}
	\mathcal{B}_{UL}(\kpmunu_h) &= \frac{n_{UL}}{N_K\cdot A(m_h)}\,, \\
	\abs{U_{\mu 4}}^2 &= \frac{\br{\kpmunu_h}}{\br{\kpmunu_\mu}} \times \frac{1}{f(m_h)}\,,
\end{alignat*}
where \(A(m_h)\) and \(f(m_h)\) are the acceptance and phase space factors for a heavy neutrino of mass \(m_h\).
The final limits are shown in \reffig{u2}.

\begin{figure}[htb]
	\centering
	\subfloat[\labelfig{mu}]{\includegraphics[width=.48\textwidth]{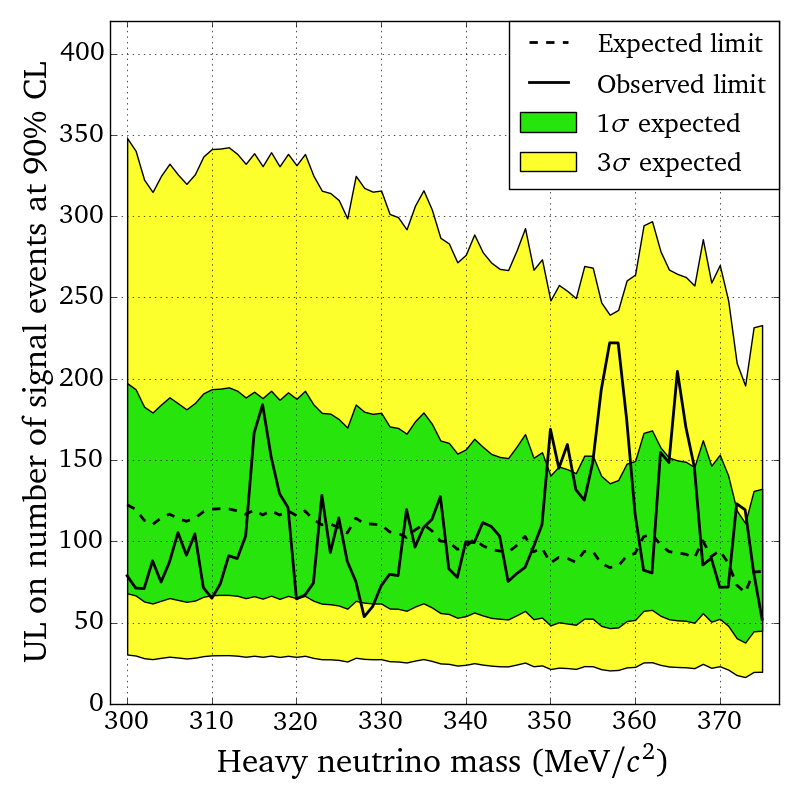}}
	\subfloat[\labelfig{u2}]{\includegraphics[width=.5\textwidth]{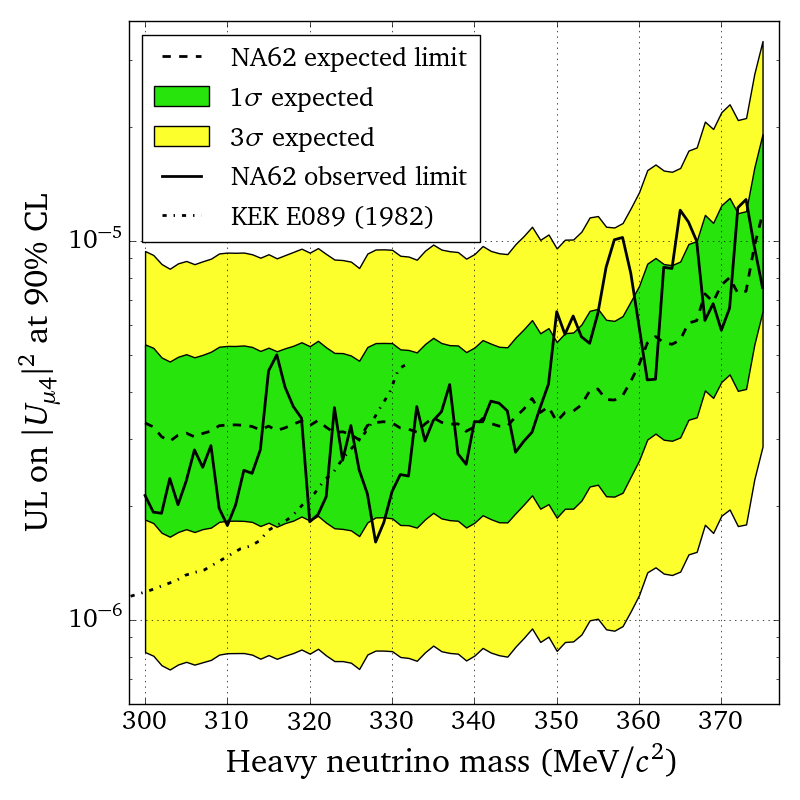}}
	\caption{
		(a) Expected and observed upper limits (at \npcl) on the number of \(\kpmunu_h\) events at each tested mass.
		(b) Expected and observed upper limits (at \npcl) on the mixing matrix element \(\abs{U_{\mu 4}}^2\) at each tested mass.
		The existing limit from KEK E089~\protect\cite{hayano_1982} is also shown.
		Below \SI{300}{\mevpercsq} there is a limit of \(\mathcal{O}(\num{e-8})\) from BNL E979~\protect\cite{artamonov_2015}.
	}
\end{figure}

\section{Status of the NA62 experiment}
\subsection{The \texorpdfstring{\(\kpinunu\)}{K+pi+nunu} decay}
The \(\kppinunu\) decay is a flavour changing neutral current process proceeding through box and electroweak penguin diagrams.
It is strongly suppressed by a quadratic GIM mechanism.
The dominant contribution comes from the short distance physics of the top quark loop, with a small charm quark contribution and long-distance corrections.
The hadronic matrix elements are extracted from the well-known semi-leptonic decay \(K^+\to\pi^0e^+\nu\). 
The standard model prediction is computed to a high degree of precision, which is limited by the uncertainties on the CKM parameters~\cite{buras_2015}:
\begin{equation*}
	\br{\kppinunu}_\text{SM} = \num{9.11(72)e-11}\,.
\end{equation*}
This decay channel is extremely sensitive to physics beyond the SM, probing mass scales beyond those accessible from direct searches at the LHC.
The current experimental determination of the branching ratio comes from 7 candidates events detected at the E787/E949 experiments~\cite{artamonov_2008}:
\begin{equation*}
	\br{\kppinunu}_\text{exp} = 17.3^{+11.5}_{-10.5}\times 10^{-11}\,.
\end{equation*}
A measurement with a precision at the level of \perc{10} would provide strong constraints on new physics scenarios.

\subsection{The NA62 detector}
The fixed target NA62 experiment aims at measuring the \(\kppinunu\) decay with \perc{10} precision.
A sample of about \num{e13} kaon decays should be collected in few years of data-taking using the \SI{400}{\gevperc} primary SPS proton beam.
A maximum of \perc{10} of background contamination is required, necessitating a background rejection factor of the order of \num{e12}.
The beam impinges on a beryllium target producing secondary particles, of which \perc{6} are kaons.
A \SI{100}{\m} long beam line selects, collimates, focuses and transports charged particles of \SI{75\pm0.8}{\gevperc} momentum to the evacuated fiducial decay volume.

\reffig{na62-detector} shows the experimental apparatus in operation since 2014.
The KTAG is a differential Cherenkov detector filled with \(N_2\) placed in the beam to identify and timestamp kaons.
It is followed by the Gigatracker (GTK), three silicon pixel stations of \SI{6x3}{\cm} surface exposed to the full \SI{750}{\mega\hertz} beam rate.
It is used to timestamp and measure the momentum of the beam particles before entering the vacuum region downstream.
The CHANTI detector placed after the Gigatracker tags hadronic interactions in the last GTK station.
The magnetic spectrometer made of four straw chambers and a dipole magnet between the second and third chamber is used to measure the momentum of downstream charged particles.
It is followed by a \SI{17}{\m} long RICH counter filled with Ne, used to separate \(\pi^+, \mu^+\) and \(e^+\).
The time of charged particles is measured both with the RICH and with an array of scintillator (CHOD) located downstream of the RICH.
Two hadronic calorimeters (MUV1 and MUV2) and a fast scintillator array (MUV3) provide further separation between \(\pi^+\) and \(\mu^+\).
A set of photons vetoes (LAVs, LKr, IRC, SAC) hermetically cover angles up to \SI{50}{\milli\rad} for electromagnetic activity.
A detailed description of the apparatus and its performances in 2015 can be found in Cortina Gil~\cite{cortina_2017}.

\begin{figure}
	\centering
	\includegraphics[width=\textwidth]{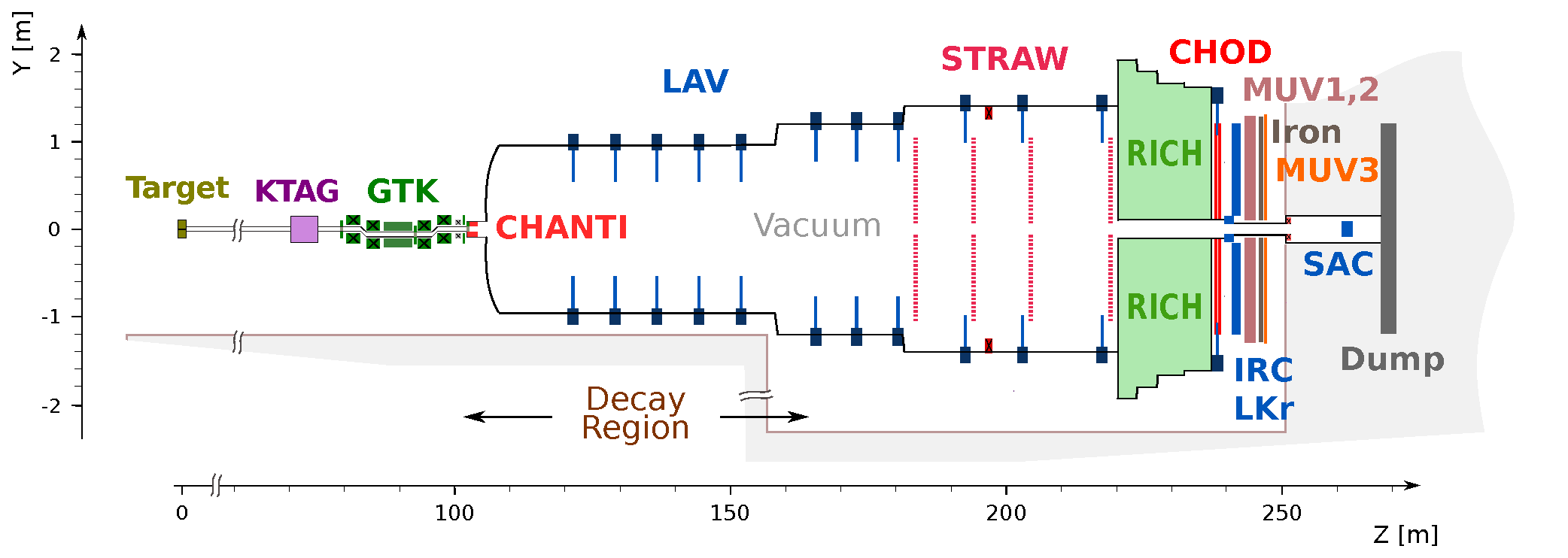}
	\caption{
		Schematic of the NA62 layout.
	}
	\labelfig{na62-detector}
\end{figure}

The detector has been completed and is fully commissioned, as well as the beam line and the high level trigger.
Two data samples have been collected: a minimum bias sample at \perc{1} intensity in 2015, used for the detector performances analysis described here; 
The second sample is used for \(\kppinunu\) analysis and was taken in stable conditions at \perc{40} of the nominal intensity in 2016.

\subsection{Detector performances}
The data taken in 2015 are used to measure the detector performances.
The timing, kinematic resolution, particle identification and photon rejection are verified against the design expectations.
A selection similar to the \(\kppinunu\) one is used for this purpose.

Tracks reconstructed in the spectrometer are selected, and a match in time and space is requested with a CHOD signal, energy deposition in the calorimeters, and a GTK track.
A track not forming a common vertex within the decay region, bound by the last GTK station and the first Straw station, with any other in-time track defines a single-track event.
To select kaon decay events, the reconstructed vertex is required to be within the \SI{65}{\m} fiducial decay region, and to be in-time with a kaon signal in the KTAG.
\reffig{mmiss-kaon} shows the \(m^2_\text{miss} = \group{P_K-P_{\pi^+}}^2\) versus \(P_{\pi^+}\) distribution for single-track events compatible with a kaon decay, where \(P_K\) is the reconstructed four-momentum of the beam kaon from the GTK track, and \(P_{\pi^+}\) is the reconstructed charged particle four-momentum from the spectrometer track, under the pion hypothesis.
The measured time resolution of the detectors are close to the design ones: \SI{100}{\ps} for the KTAG and below \SI{200}{\ps} for the GTK and the CHOD.

\begin{figure}
	\centering
	\subfloat[\labelfig{mmiss-kaon}]{\includegraphics[width=.5\textwidth]{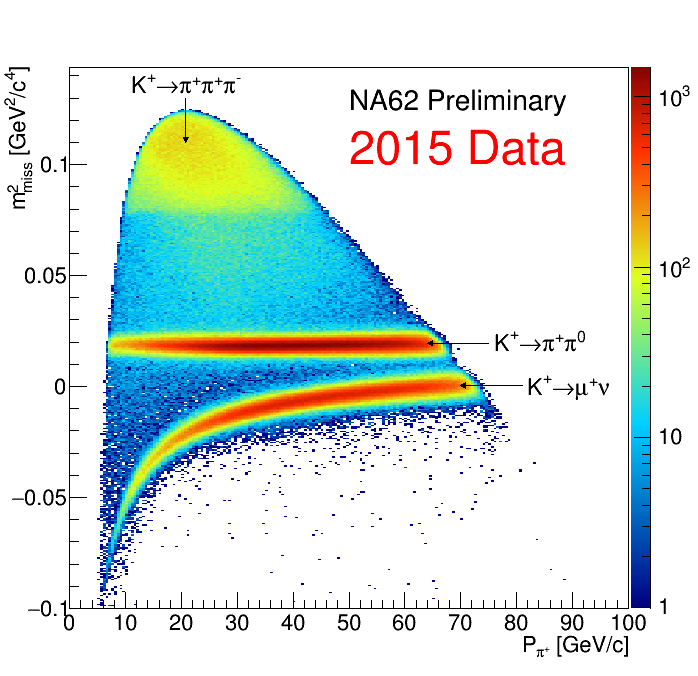}}
	\subfloat[\labelfig{pid-efficiency}]{\includegraphics[width=.5\textwidth]{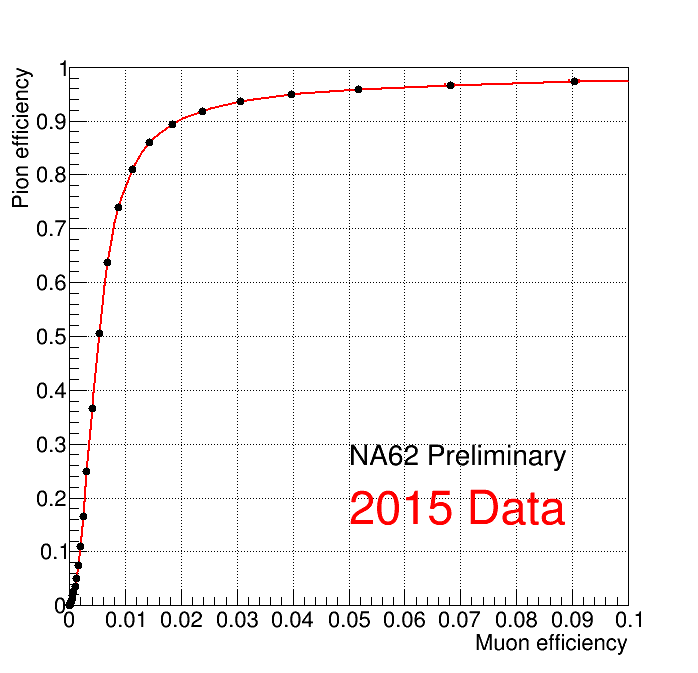}}
	\caption{
		(a) \(m^2_\text{miss}\) distribution under \(\pi+\) mass hypothesis as a function of the reconstructed track momentum for single-track kaon events.
		(b) \(\pi^+/\mu^+\) identification efficiency with the RICH detector. 
	}
\end{figure}

The resolution on \(m^2_\text{miss}\) is about \SI{1.2e-3}{\gev\squared\per\clight\tothe{4}}, close to the \SI{1e-3}{\gev\squared\per\clight\tothe{4}} design value.
The \(m^2_\text{miss}\) kinematic variable should provide a rejection factor of the order of \(\mathcal{O}\group{\numrange{e4}{e5}}\) for the main kaon decay modes by defining the signal regions (I) \(0 < m^2_\text{miss} < \SI{0.01}{\gev\squared\per\clight\tothe{4}}\) and (II) \(0.026 < m^2_\text{miss} < \SI{0.068}{\gev\squared\per\clight\tothe{4}}\).
A pure sample of \(K^+\to\pi^+\pi^0\) selected by requiring the additional presence of two photons compatible with a \(\pi^0\) in the LKr calorimeter is used to measure the kinematic rejection factor, which is found to be of the order of \num{e3}. 
The last GTK station installed in 2016 is expected to bring the performances to the design goal.

An additional \num{e7} rejection factor is provided by particle identification using the RICH detector and calorimeters.
The pion momentum is requested to be in the range \SIrange{15}{35}{\gevperc}, where the RICH is designed to achieve the best \(\pi^+/\mu^+/e^+\) separation.
The \(K^+\to\pi^+\pi^0\) sample used for kinematic studies, and a sample of \(K^+\to\mu^+\nu\) selected by requiring an additional matching signal in the MUV3 are used to study the \(\pi^+/\mu^+\) separation in the RICH.
\reffig{pid-efficiency} shows a \num{e2} \(\mu^+\) rejection factor for \perc{80} \(\pi^+\) efficiency, reaching \perc{90} in 2016 data thanks to the improved RICH mirror alignment.
A simple additional cut and count analysis involving the calorimeters increases the rejection factor to \numrange{e4}{e6} for a \(\pi^+\) efficiency in the range \SIrange{90}{40}{\%}.

The final rejection of \num{e8} for kaon decay channels involving a \(\pi^0\) is provided by the detection of at least one of the \(\gamma\) in the photon vetoes.
The high energy of the \(\pi^0\) in the \(K^+\to\pi^+\pi^0\) decay (\(E_{\pi^0}>\SI{40}{\gev}\)) and the angular correlation between the two \(\gamma\) of the \(\pi^0\) decay is exploited
A sample of \(K^+\to\pi^+\pi^0\) selected through kinematics only is used to assess the achieved rejection factor.
The measurement with 2015 data is statistically limited at \num{e6} (\npcl) as an upper limit.

The analysis of 2015 data shows that the performances are reached, or are close to, the design values for the different sub-systems.
The improvements made in preparation of the 2016 data-taking should bring the remaining performances to their design expectations.  


\section{Conclusions}
Using the data collected in 2007, a peak search in the mass spectrum of the \(\kpmunu_\nu\) decay has been performed in steps of \SI{1}{\mevpercsq}. 
New limits on the heavy neutrino production are set at the level of \SIrange{e-5}{e-6} on the mixing matrix element \(\abs{U_{\mu 4}}^2\) in the range \SIrange{300}{375}{\mevpercsq}.

The NA62 experiment has been fully commissioned, and data have been taken in 2015--2016 at \perc{1} and \perc{40} of the nominal intensity.
The study of the detector performances shows that they are in-line with the design values.
The \(\kppinunu\) analysis with the 2016 data is ongoing, and a wider physics program for short and medium term has been established.

\end{document}

%% file: cfg_commands.tex
\newcommand{\refsec}[1]{\autoref{sec:#1}}
\newcommand{\reffig}[1]{\autoref{fig:#1}}
\newcommand{\refapp}[1]{Appendix~\ref{app:#1}}
\newcommand{\refeq}[1]{eq.~\eqref{eq:#1}}
\newcommand{\reftab}[1]{Table~\ref{tab:#1}}

\newcommand{\labelsec}[1]{\label{sec:#1}}
\newcommand{\labelfig}[1]{\label{fig:#1}}
\newcommand{\labelapp}[1]{\label{app:#1}}
\newcommand{\labeleq}[1]{\label{eq:#1}}
\newcommand{\labeltab}[1]{\label{tab:#1}}

\renewcommand{\matrix}[1]{\begin{pmatrix}#1\end{pmatrix}} 
\newcommand{\pd}[2]{\frac{\partial#1}{\partial#2}} 
\newcommand{\ppd}[2]{\frac{\partial^2#1}{\partial#2^2}} 
\newcommand{\abs}[1]{\left\lvert#1\right\rvert}
\newcommand{\group}[1]{\left(#1\right)}
\newcommand{\perc}[1]{\SI{#1}{\percent}}

\newcommand{\nark}{\(\text{NA62}\text{-}{\text{R}_\text{K}}\)\xspace}
\newcommand{\narkt}{\text{NA62}\text{-}{\text{R}_\text{K}}}
\newcommand{\pizero}{{\pi^{0}}}
\newcommand{\mff}{\mathcal{F}}
\newcommand{\br}[1]{\mathcal{B}(#1)}
\newcommand{\npcl}{\SI{90}{\percent} CL\xspace}
\newcommand{\compact}{COmPACT}
\newcommand{\vckm}{V_\text{CKM}}
\newcommand{\SIrangetext}[3]{\SIrange[range-phrase = { and },range-units=repeat]{#1}{#2}{#3}}
\newcommand{\SIrangedash}[3]{\SIrange[range-phrase = {--},range-units=repeat]{#1}{#2}{#3}}
\newcommand{\state}[1]{\textbf{#1}}
\newcommand{\cmd}[1]{\textbf{#1}}
\newcommand{\svc}[1]{\textit{#1}}

\newcommand{\kpinunu}{K^\pm\to\pi^\pm\nu\bar{\nu}}
\newcommand{\kppinunu}{K^+\to\pi^+\nu\bar{\nu}}
\newcommand{\knpinunu}{K_L\to\pi^0\nu\bar{\nu}}
\newcommand{\pinunu}{K\to\pi\nu\bar{\nu}}
\newcommand{\kpp}{K_{2\pi}}
\newcommand{\kpipi}{K^\pm\to\pi^\pm\pi^0}
\newcommand{\kpmn}{K_{\mu3}}
\newcommand{\kpimunu}{K^\pm\to\pi^0\mu^\pm\nu}
\newcommand{\kmn}{K_{\mu2}}
\newcommand{\kmunu}{K^\pm\to\mu^\pm\nu}
\newcommand{\kpen} {K_{e3}}
\newcommand{\kpienu} {K^\pm\to\pi^0 e^\pm\nu}
\newcommand{\ken} {K_{e2}}
\newcommand{\kenu} {K^\pm\to e^\pm\nu}
\newcommand{\kpipipi}{K^\pm\to\pi^\pm\pi^+\pi^-}
\newcommand{\kppp}{K_{3\pi}}

\newcommand{\pizerod}{\pi^{0}_{D}}
\newcommand{\pizerodalitz}{\pi^{0}\to e^+e^-\gamma}
\newcommand{\kppd}{K_{2\pi D}}
\newcommand{\kpipid}{K^\pm\to\pi^\pm\pi^0_D}
\newcommand{\kpmnd}{K_{\mu3 D}}
\newcommand{\kpimunud}{K^\pm\to\pi^0_D\mu^\pm\nu}
\newcommand{\kpend} {K_{e3D}}
\newcommand{\kpienud} {K^\pm\to\pi^0_D e^\pm\nu}

\newcommand{\figwidth}{0.7\columnwidth}
\newcommand{\figwidthcol}{0.5\linewidth}

\newcommand{\cmark}{\ding{51}}
\newcommand{\xmark}{\ding{55}}

\newcommand{\figscan}[2]{
\begin{figure}
	\center
	\subfloat[\labelfig{scan_#1_fit}]{\includegraphics[width=\figwidthcol]{scan_#1_fit}}
	\subfloat[\labelfig{scan_#1_sel}]{\includegraphics[width=\figwidthcol]{scan_#1_sel}}
	\caption{#2}
	\labelfig{scan_#1}
\end{figure}
}